\documentclass[prl,twocolumn,showpacs,amsmath,amssymb]{revtex4}

\usepackage[dvips]{graphicx}
\usepackage{graphics}
\usepackage{dcolumn}
\usepackage{bm}
\usepackage[usenames]{color}

\newcommand{\ve}{\mathbf}

\begin{document}

\title{\begin{center} Magnetization-Switched Metal-Insulator
Transition in a (Ga,Mn)As Tunnel Device
 \end{center}}

\author{K. Pappert, M.J. Schmidt, S. H\"{u}mpfner, C. R\"{u}ster, G.M. Schott, K. Brunner,
C. Gould, G. Schmidt, and L.W. Molenkamp}

\affiliation{Physikalisches Institut (EP3), Universit\"{a}t
W\"{u}rzburg, Am Hubland, D-97074 W\"{u}rzburg, Germany}

\date{\today}

\begin{abstract}
We observe the occurrence of an Efros-Shklovskii gap in (Ga,Mn)As
based tunnel junctions. The occurrence of the gap is controlled by
the extent of the hole wave-function on the Mn acceptor atoms. Using
$\bf{k} \cdot \bf{p}$-type calculations we show that this extent
depends crucially on the direction of the magnetization in the
(Ga,Mn)As (which has two almost equivalent easy axes). This implies
one can reversibly tune the system into the insulating or metallic
state by changing the magnetization.
\end{abstract}


\maketitle

A very direct way to observe the Efros-Shklovskii (ES) gap, the soft
gap induced by Coulomb correlations near the Fermi level of a Mott
insulator\cite{Efros,Shklovskii}, is by means of tunnel
spectroscopy. Such experiments were, e.g., performed on the
(three-dimensional) nonmetallic doped semiconductor Si:B
\cite{Massey,Lee} and on thin (two-dimensional) Be
films\cite{Butko}. While both of these experiments employed large
area tunnel junctions and a metallic counter electrode, a more
recent study employed Ge:As break junctions\cite{Sandow}. This
latter approach avoids possible screening of the Coulomb
correlations, but the mesoscopic character of the contact may
complicate extraction of bulk Coulomb gap behaviour\cite{Larkin}.

We have recently investigated the physics of a novel type of
magnetoresistance, dubbed TAMR (tunneling anisotropic
magnetoresistance) \cite{Chris1,Chris2}. TAMR results from the
dependence of the density of states (DOS) in strongly spin-orbit
coupled ferromagnetic semiconductors, such as (Ga,Mn)As, on the
direction of the magnetization of the material. In \cite{Chris2} we
reported a drastic ($> 10^4$) increase of the spin-valve signal in a
(Ga,Mn)As/GaAs/(Ga,Mn)As tunnel structure on lowering the sample
temperature from 4.2 to 1.7 K, and speculated that this behaviour
might result from the opening of an ES gap. Here, we provide
evidence that the high resistance state of the sample indeed
corresponds to a soft-gapped Mott insulator. In these samples, the
metal-to-insulator transition (MIT) is driven by a large variation
of the Bohr radius of a hole bound to a Mn-impurity when the
magnetization of the layer is switched from one easy axis to the
other. This assignment is supported by a $\bf{k} \cdot \bf{p}$-type
calculation of a hydrogen-like impurity in a ferromagnetic GaAs
host, extending the successful mean field model for (Ga,Mn)As
\cite{Abolfath, Dietl}.

\begin{figure}[h!]
\includegraphics[angle=0,width=8cm]{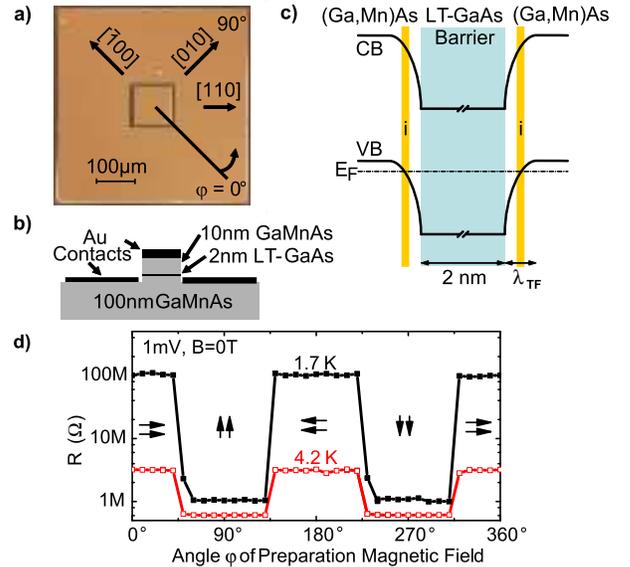}
\caption{a) Optical micrograph of the sample. b) Sample layer
structure. c) schematic band diagram illustrating the barrier for
hole transport, the depletion region near the interfaces, and the
effective injector layers (i). d) 4.2 K and 1.7 K magnetoresistance
in zero field at 1 mV bias after preparing the sample by magnetizing
in various field directions.}. \label{figure1} \vspace*{-0.5 cm}
\end{figure}

Our (Ga,Mn)As tunnel structure is shown in Fig.~\ref{figure1}b. From
bottom to top, the Ga$_{0.94}$Mn$_{0.06}$As (100~nm)/ GaAs (2~nm)
/Ga$_{0.94}$Mn$_{0.06}$As (10~nm) trilayer stack has been grown by
low temperature molecular beam epitaxy (LT-MBE) on a semi-insulating
GaAs substrate and a 120 nm undoped GaAs buffer layer. Both
(Ga,Mn)As layers are ferromagnetic with an as-grown Curie
temperature of $\sim65$ K and highly p-type due to the intrinsic
doping arising from the Mn atoms.

As seen in the optical micrograph of Fig.~\ref{figure1}a, the layer
stack is patterned into a square mesa of 100$\times$100~$\mu$m$^2$
by positive optical lithography, metal evaporation, lift-off and wet
etching. The top contact is in-situ Ti/Au. Contact to the lower
(Ga,Mn)As layer is established by a W/Au ring around the mesa,
fabricated by metal deposition and optical lithography
\cite{Chris2}.

The structure is intended for vertical tunneling transport
experiments. A key element in this device design is that the LT-GaAs
tunnel barrier contains many defects, and thus acts as an effective
carrier sink, resulting in a thin depleted (Ga,Mn)As region near the
barrier and thus band bending near the interfaces, as sketched in
Fig.~\ref{figure1}c.

We measure the resistance of the device depending on its magnetic
history. Using a magnetocryostat equipped with three pairs of
Helmholtz coils, we magnetize the sample along a specific angle
$\varphi$ in the sample plane. One of the magnetic easy axes of the
(Ga,Mn)As layers, defined as $\varphi=0^{\circ}$, is along the [100]
crystal direction. The second, perpendicular, easy axis is along
[010] corresponding to $\varphi=90^{\circ}$. As the field is swept
to zero, the magnetization relaxes to the nearest easy axis, and
this magnetization state determines the resistance.
Fig.~\ref{figure1}d illustrates the results of such measurement
\cite{amp} for many preparation angles $\varphi$. The 4.2 K-curve
can be understood as resulting from TAMR \cite{Chris1}: the
anisotropy in the momentum dependent DOS with respect to
magnetization direction ([100] or [010]) causes a resistance
difference between these two magnetization orientations.

\begin{figure}[h!]
\vspace*{-0 cm}
\includegraphics[angle=0,width=8cm]{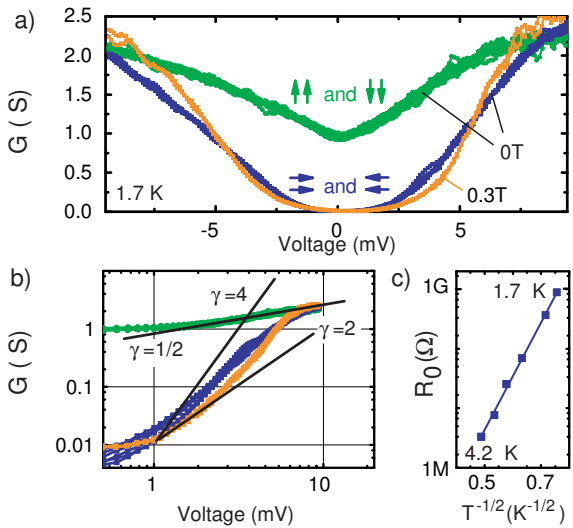}
\vspace*{-.1 cm} \caption{a)Differential conductance-voltage curves
at 1.7 K including sets of curves belonging to the two magnetization
states at B=0, and one set at B=300 mT along $\varphi$ = 3 and
6$^{\circ}$. b)Log-log plot of same data. c) Temperature dependence
of the zero bias resistance of the high resistance state.}
\label{figure2} \vspace*{-0.2 cm}
\end{figure}

While TAMR is a band structure effect, the extreme amplification of
the effect between 4.2 and 1.7 K can only be understood by
considering many-particle effects. The data of Fig.~\ref{figure2}a
and b further support the presence of electron-electron
correlations.

The differential conductance $G(V)=dI/dV$ in Fig.~\ref{figure2}b is
calculated from current-voltage-characteristics taken in the
remanent magnetization state after preparation of the sample at an
angle $\varphi$. In Fig.~\ref{figure2}a, the $G(V)$ curves at 1.7 K
are segregated into two groups associated with each of the two
resistance states identified in Fig.~\ref{figure1}d. The
magnetization of both (Ga,Mn)As layers relaxes into the low
resistance state corresponding to magnetization along [$010$] for
all preparation angles between 45$^{\circ}$ and 135$^{\circ}$(upper
curves in Fig.~\ref{figure2}a). It relaxes along the
[100]-high-resistance state for preparation fields within
45$^{\circ}$ of the [100] direction (lower curves).

The $G(V)$ characteristics have a distinct shape for each
magnetization direction. High conductance curves are
square-root-like and show non-vanishing conductance at zero voltage.
This is typical for the metallic behaviour of doped semiconductors
(here (Ga,Mn)As) taking part in the tunnel process. The square root
dependence stems from weak localization in metals \cite{Altshuler}.
The high resistance state, however, shows insulating behaviour. The
conductance vanishes at zero voltage and the curves follow a higher
power law as seen in the log-log plot of Fig.~\ref{figure2}b.

The shape of these $G(V)$ curves is reminiscent of those observed by
Lee et al. \cite{Lee} in their investigations of tunneling from a
metal into disordered boron-doped silicon near the MIT. Their theory
describes the formation of a soft ES-gap \cite{Efros} stemming from
an increase in electron-electron Coulomb interaction as screening is
reduced when the MIT is approached from the metallic side. This gap
around the Fermi energy is observable in the low bias tunneling
conductance. It manifests itself in a different power law behaviour
$G(V)\propto V^m$, with $m=1/2$ for metallic (just above the MIT)
and $m>2$ for insulating (below the MIT) material.

The transition from metallic to insulating behaviour is thermally
activated as demonstrated by the zero bias differential resistance
data taken at an angle corresponding to the high resistance state at
1.7 K, shown in Fig.~\ref{figure2}c. The plot shows that the data is
consistent with an exponential activation following $R\propto
exp(1/T)^{1/2}$ expected for an ES gap material under the usual
assumption of single hop tunneling \cite{Sandow, Efros}. We note
however that because the accessible temperature range spans less
then one decade, and we cannot rule out an $R\propto exp(1/T)^{1/4}$
dependence expected from a non correlated Mott transition.

\begin{figure}[h!]
\includegraphics[angle=0,width=8cm]{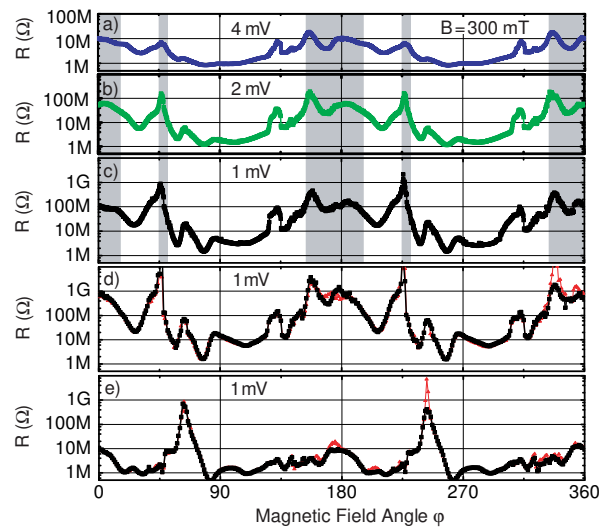}
\caption{Resistance of the sample at 1.7 K and 300 mT as a function
of magnetic field direction under various bias and on different
cooldowns. d) and e) each show the results of two separate
measurement, confirming reproducibility.} \label{figure3}
\vspace*{-.2 cm}
\end{figure}

A further clue to the origin of our effect comes from the data of
Fig.~\ref{figure3} showing the resistance behaviour of the sample at
300 mT, a field sufficient to force the magnetization vectors
parallel to the field direction for all field angles. The figure
shows very strong and apparently random oscillations reminiscent of
quantum interference effects, and likely arising from a
statistically defined electronic state in the sample.
Fig.~\ref{figure3} a-c show that the amplitude of these oscillations
grows significantly with reduced bias for angular ranges (gray
regions in the figure) associated with ES behaviour.
Fig.~\ref{figure3}c-e show a significant effect upon thermal cycling
to temperatures of some tens of K (above the ES activation
temperature). The three curves are nominally identical measurements
after subsequent thermal cycling. The main behaviour of the device
remains unchanged, but the details of the fluctuations change
significantly, presumably corresponding to a new impurity
configuration. We emphasize that the change in the fluctuation
pattern is due to thermal cycling. On a given cooldown, measurements
are reproducible as confirmed by the two, almost indistinguishable,
sets of measurements included in each of Fig.~\ref{figure3}d and e.
$G(V)$ curves at 300 mT along the $\varphi$ = 3 and 6$^{\circ}$
given in Fig.~\ref{figure2}c show similar behaviour to the other
high resistance curves.

We attribute these fluctuations to quantum interference effects on
the variable range hopping that is the transport mechanism in the
insulating state \cite{Raith}. In reference \cite{Hughes}, such
quantum interference was confirmed by a statistical analysis of the
amplitude of the fluctuations as the Fermi energy is varied using an
external gate bias. A similar analysis cannot be reliably applied
here as the relation between the magnetization direction and the
Fermi energy is complex and non-linear.

The behaviour of our sample is fully consistent with the depleted
(Ga,Mn)As injection layer undergoing a MIT triggered by a
reorientation of the magnetization from [010] to [100], and the
formation of an ES-gap. In order to understand how a change in
magnetization direction can trigger a MIT, we need to consider the
criterion for the passage from metallic to insulating properties.

Metallic transport properties require the charge carriers to be in
long range itinerant states. This condition implies significant
overlap of the wavefunctions between acceptors (Mn atoms). As such,
the passage through the MIT depends not only on the carrier density,
but also on the volume occupied by the wavefunctions of the states
of the individual dopants. Indeed a change in impurity conductivity
associated with a strain induced change in Bohr radius has been
reported previously \cite{Pollak}.

For (Ga,Mn)As, it is well established \cite{Abolfath, Dietl} that
details of the DOS are influenced by the magnetization direction. To
determine if this plays a role in renormalizing the wavefunction
extent of our localized states, we turn to a numerical calculation
of bound hole states.

The valence band in a zinc blende semiconductors is described in a
6-band $\bf{k}\cdot\bf{p}$ framework by the full Luttinger
Hamiltonian $H(\ve k)$ \cite{LongDietl}. Luttinger and Kohn showed
\cite{kohnluttinger1} that a charged impurity in a semiconductor
host can be described by treating the impurity wavefunction as an
envelope wavefunction $\ve B(\ve k)$ of Bloch waves. The
time-independent Schr\"{o}dinger equation is then
\cite{kohnluttinger1}:

\vspace*{-.2 cm}
\begin{equation}
H(\ve k)\ve B(\ve k) +\int d^3\ve k'\mathcal U(\ve k-\ve k')\ve
B(\ve k')=E_b \ve B(\ve k)\label{origkohnluttinerimpurityequation}
\end{equation}

where $E_b$ is the binding energy. $\mathcal U$ is the Fourier
transform of the Coulomb limit of the Yukawa interaction potential
between the hole and impurity-center. $H(\ve k)$ is a
$6\times6$-matrix acting on the 6 dimensional vector $\ve B(\ve k)$
locally in $\ve k$-space. It describes the bandstructure including,
for magnetic (Ga,Mn)As, pd-exchange, strain and SO-coupling.
Eq.(\ref{origkohnluttinerimpurityequation}) is solved numerically
for the acceptor ground state in a magnetic zinc blende
semiconductor. The obtained k-space wavefunctions are fitted to
hydrogen ground state wavefunctions in order to extract an effective
Bohr radius for each k direction and for various magnetization
directions. Details of these calculations will be published
elsewhere.

\begin{figure}
\centering
\includegraphics[width=8cm]{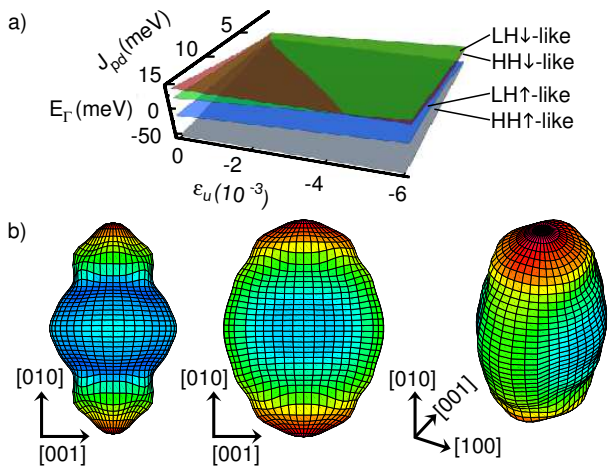}
\caption{a) $\Gamma$ point energy of the four top bands as a
function of strain $\epsilon_{u}$ and pd-exchange. b) Extents of a
hole bound to a Mn-impurity. [001] is the growth direction and
$\epsilon_{u}$ is along [010]. Left: Magnetization ($\mathbf
M\parallel [100]$). Middle:($\mathbf M
\parallel [010]$). Right: Perspective view for $\mathbf M\parallel
[010]$.} \label{theoryfigure} \vspace*{-.5 cm}
\end{figure}

From symmetry considerations, one expects that our (Ga,Mn)As has
four-fold in-plane magnetic easy axes. Experimentally however, a
small additional [010] uniaxial anisotropy is generally observed in
samples from various sources. While the origin of this uniaxial
anisotropy is unknown, considerable insight is gained by
phenomenologically introducing a uniaxial strain term $\epsilon_{u}$
along [010] \cite{Chris1}. An important result of the bulk band
structure calculation is that the energy ordering of the bands
depends on the relative strain and magnetization directions. In
Fig.~\ref{theoryfigure}a, the $\Gamma$ point energy of the top 4
valance bands is plotted as a function of the amplitude of
$\epsilon_{u}$ and the pd- exchange. For zero strain, we obtain the
expected \cite{Abolfath,Dietl} nearly equidistant spacing between
the 4 bands, while increasing strain shifts the relative energy
position of the light-hole-like bands with respect to the
heavy-hole-like bands, leading to a change in ordering of the bands
at the line defining the crossing point between the top two planes
in the figure, and thus to a profound change in the character of the
wavefunction at a given energy. This is directly reflected in the
wavefunction of the impurity hole state.

Fig.~\ref{theoryfigure}b shows the extent of a bound hole for
magnetization along different in-plane easy axes with an amplitude
of $\epsilon_{u}$ beyond the crossing point for the bulk energy at
the $\Gamma$-point. There is a significant difference in the size of
the wavefunction between the $\vec M||\epsilon_{u}$ and the $\vec M
\perp \epsilon_{u}$-state. The result is that the wave function
overlap between neighboring dopant sites depends on the direction of
the magnetization, and for reasonable parameters is consistent with
the observation of the magnetization reorientation induced MIT.

It is important to note that the above calculation does not predict
a MIT for the bulk (Ga,Mn)As which, as typical in high quality
(Ga,Mn)As, has nearly temperature independent resistivity in the
temperature range of interest. Our prediction is limited to the thin
(Ga,Mn)As layer near the interface with the tunnel barrier. The
barrier consists of LT-GaAs, a material with many mid-gap traps and
leading to a gradual spatial depletion of the (Ga,Mn)As near the
barrier on the length scale of the Thomas-Fermi screening length of
$\sim 2 {\AA}$, and thus to a much lower effective local carrier
density, as illustrated schematically in Fig.~\ref{figure1}c. If the
thin depleted layer is close enough to the MIT, a change in the
wavefunction extent can trigger the transition. Another point is
that the mean-field p-d exchange strength is smaller for lower hole
densities. As seen in Fig.~\ref{theoryfigure}, a weaker pd-exchange
lowers the amplitude of $\epsilon_{u}$ needed to reach the band
crossing to much smaller values that are consistent with those used
in the original description of TAMR \cite{Chris1,Chris2}.

A key point here is that while this thin depleted layer plays a
limited role in the magnetic properties of the (Ga,Mn)As, it has a
dominant effect on the transport. The reason for this lies in the
very short mean free path of holes in (Ga,Mn)As which is of the
order of a few ${\AA}$. Combining the fact that, by definition, the
transition from diffusive to tunneling transport takes place at a
density where holes can no longer classically diffuse, with this
extremely short mean free path, must lead to a very thin effective
injector layer from which the tunneling originates. This injection
layer is characterized by considerably reduced carrier density
relative to the bulk, and thus consistent with our model description
above.

The link between the MIT and the development of an ES-gap is that
the reduction in mobile carriers reduces screening. With unscreened
Coulomb-interactions between localized states, the DOS at the
Fermi-energy for carrier concentrations just below the MIT vanishes
with a dimension-dependent power law\cite{Efros}, i.e. $D(\epsilon)
\propto \epsilon$, where $\epsilon$ is the energy measured with
respect to the Fermi energy, for two-dimensional materials, and
$D(\epsilon) \propto \epsilon^2$ in 3D. Of course, we cannot rule
out that the change in screening causes a slight shift in position
of the effective injector thus modifying the tunneling distance, and
accounting for part of the observed resistance change.

Larkin and Shklovskii\cite{Larkin} derived a power law behaviour for
conductance versus voltage (G-V) curves for tunneling between two
three dimensional localized materials where the parabolic DOS
$D(\epsilon) \propto \epsilon^2$ on each side multiply and lead to
$G\propto V^6$. In our case, given the very thin nature of the the
injector, a 2-D description of the DOS of the tunneling reservoirs
is likely more appropriate. Thus, the DOS is linear in energy, and
the expected power of the G-V curves is 4. Moreover, it is unclear
in our experiment whether both sides of the barrier play an
equivalent role, or whether the effects are dominated by one
electrode. Depending on the relative role of the two barriers, one
would thus expect a power law somewhere between $\propto V^2$ and
$\propto V^6$. Neglecting very low voltages where thermal smearing
is important, this prediction is in good agreement with the
experimental observation of Fig.~\ref{figure2}.

In conclusion, we have observed the magnetization reorientation
induced MIT in a (Ga,Mn)As based transport sample. The transition
can be understood as stemming from a modification of the wave
functions of individual dopants due to the coupling of the Mn
dopants to the magnetization direction in the bulk. Furthermore, the
transition is accompanied by the opening of an ES-gap at the Fermi
energy which manifests itself as a change in the power law behaviour
in conductance-voltage characteristics in tunneling experiments.
This is the first observation of a MIT induced by a change in
magnetization \emph{direction} in any material. In addition to the
fundamental interest intrinsic to these observations, the results
may also have technological relevance in opening up new
possibilities of controlling the transport properties of devices by
magnetization reversal.

The authors thank R. Opperman, M. Flatt\'e , R. Giraud, and B.I.
Shklovskii for useful discussions and V. Hock for sample
preparation. We acknowledge financial support from the EU (SPINOSA
and FP6-IST-015728, NANOSPIN), the German BMBF(13N8284) and DFG
(BR1960/2-2).


\begin{thebibliography}{99}

\bibitem{Efros} A.L. Efros and B.I. Shklovskii, J. Phys. C {\bf 8}, L49
(1975).

\bibitem{Shklovskii} B.I. Shklovskii and A.L. Efros, {\it Electronic
Properties of Doped Semiconductors} (Springer, New York, 1984).

\bibitem{Massey} J.G. Massey and M. Lee, Phys. Rev. Lett. {\bf 75},
4266 (1995).

\bibitem{Lee} M. Lee, J.G. Massey, V.L. Nguyen and B.I. Shklovskii,
Phys. Rev. {\bf B60}, 1582 (1999).

\bibitem{Butko} V.Yu. Butko, J.F. diTusa, and P.W. Adams, Phys. Rev.
Lett. {\bf 84}, 1543 (2000).

\bibitem{Sandow} B. Sandow, K. Gloos, R. Rentzsch, A.N. Ionov, and
W. Schirmacher, Phys. Rev. Lett. {\bf 86}, 1845 (2001).

\bibitem{Larkin} A.I. Larkin, B.I. Shklovskii, phys. stat. sol. (b)
{\bf 230}, 189 (2002)

\bibitem{Chris1} C. Gould et al., Phys. Rev. Lett. {\bf 93}, 117203 (2004).

\bibitem{Chris2} C. R\"{u}ster et al., Phys. Rev. Lett. {\bf 94}, 027203 (2005).

\bibitem{amp} The amplification is such that the current in the high
resistant state is comparable to amplifier offsets. A constant
offset current was removed from the data before analysis. This
brings into question the accuracy of the resistance for values above
1 $G\Omega$.

\bibitem{Abolfath} M. Abolfath, T. Jungwirth, J. Brum and A.H.
MacDonald, Phys. Rev. B{\bf 63}, 054418 (2001)

\bibitem{Dietl} T. Dietl, H. Ohno, F. Matsukura, Phys. Rev. B
{\bf 63}, 195205 (2001)

\bibitem{Altshuler}B.L. Altshuler and A.G. Aronov, Sov. Phys. JETP
\textbf{50}, 968, (1979).

\bibitem{Raith} M.E. Raikh and I.M. Ruzin, Sov. Phys. JETP 65, 1273,
(1987).

\bibitem{Hughes} R.J.F. Hughes et al., Phys. Rev. B{\bf 54}, 2091
(1996).

\bibitem{Pollak} F.H. Pollak, Phys. Rev. \textbf{138}, 618, (1965)

\bibitem{LongDietl} T. Dietl, H. Ohno and F. Matsukura, Phys. Rev. B \textbf{63}, 195205

\bibitem {kohnluttinger1} J.M. Luttinger and W. Kohn, Phys. Rev. \textbf{97}, 869 (1955)






\end{thebibliography}
\end{document}